\documentclass[12pt]{article}
\usepackage{sbc-template}

\usepackage{flafter}

\usepackage[USenglish]{babel}
\usepackage{graphicx}
\usepackage{overpic}

\usepackage[T1]{fontenc}
\usepackage{newtx}

\usepackage[style=numeric]{biblatex}
\usepackage{csquotes}

\usepackage{booktabs}

\usepackage{url}

\author{Davi Antônio da Silva Santos\inst{1}, Bruno César Ribas\inst{1}}
\address{Faculdade do Gama (FGA) -- Universidade de Brasília (UnB)\\ Área Especial de Indústria Projeção A -- Setor Leste -- Gama -- 72.444-240 -- DF -- Brazil
	\email{antoniossdavi@gmail.com, bruno.ribas@unb.br}
}

\title{Maratona Linux a tale of upgrading from Ubuntu 20.04 to 22.04}

\begin{document}
	
	
	\maketitle
	
	\begin{abstract}
		Maratona Linux is the development environment used since 2016 on the ``Maratona de Programação'', ICPC's South American regional contest. It consists of Debian packages that modify a standard Ubuntu installation in order to make it suitable for the competition, installing IDEs, documentation, compilers, debuggers, interpreters, and enforcing network restrictions. The project, which began based on Ubuntu 16.04, has been successfully migrated from 20.04 to 22.04, the current Long-term Support (LTS) version. The project has also been improved by adding static analyzers, updating the package dependency map, splitting large packages, and enhancing the packaging pipeline.
	\end{abstract}
	
	\section{Introduction}
The ``Maratona de Programação'' is a competition for undergraduates and those who have begun attending post-graduation courses related to Computer Science. It is organized by the Brazilian Computing Society since 1996 \cite{sbc_maratona_22}. On account of being the South American regional contest of the International Collegiate Programming Contest (ICPC), the teams from Brazilian and South-American universities compete for a spot in the ICPC World Finals \cite{sbc_maratona_22}.

The Maratona Linux project is the standard development environment used on the ``Maratona de Programação'' since 2016 when it was still based on the Ubuntu Long-term Support version (LTS) 16.04 \cite{maratona_linux_2019}. It consists of Debian-formatted software packages that modify a standard desktop Ubuntu LTS installation, transforming it into a development environment suitable for the competition.

The modifications made by Maratona Linux enable the contestants to develop, debug, execute and submit their solutions developed in C, C++, Python 3, Java 17, and Kotlin 1.6 using free and proprietary Integrated Desktop Environments (IDEs). Compilers, debuggers, static analyzers and reference documentation are also included, as the competition is often run in environments with restricted Internet connections.

Due to improvements in the programming tools, graphical interfaces, and the changes in system requirements imposed by updates in the IDEs, Maratona Linux must be continuously updated. These updates must be compatible with the most recent Ubuntu LTS, which is made available by its developer, Canonical Ltd., every two years \cite{ubuntu_release_cycle_2022}.

This work is divided into four sections. It describes the process of migrating the Maratona Linux from its previous state on Ubuntu 20.04 to 22.04, the current LTS version, and the improvements that have been made to the project.

This paper is organized as follows: the first section, Background, details important concepts and briefly introduces Debian, Snap, and Flathub technologies. The second one, Development and Tests, details the development process. The third section, Migration Status, specifies the alterations made to enable the migration, and the improvements made to the project. The fourth and last section (Final Remarks) summarizes the article and points to possible improvements.
	
	\section{Background}
The Maratona Linux project uses multiple technologies to achieve its expected functionalities, such as the Debian (Deb)  packaging format and its pipeline. The Snap packaging format and the Flatpak framework were used previously. This section details the history of the project and explains basic concepts about the aforementioned technologies.

\subsection{Maratona Linux}
The project is a set of Debian packages that modify a standard desktop installation of Ubuntu LTS to standardize the development environment used by the contestants. The modifications range from installing the Integrated Development Environments (IDEs) recommended by the competition's organizers to restrictions on network communication and configurations to enable the submission of the solutions via the BOCA online judge \cite{maratona_linux_2019, Campos2004, Campos2010}.

Unlike the official ICPC disk image, which is used for the World Finals Contest \cite{icpc_disk_2021}, Maratona Linux does not require a custom installation disk: it can be installed from a Debian package repository on a Ubuntu LTS installation. Although a regular installation of Maratona Linux still requires a dedicated machine, as some uninstallation routines have not been implemented yet, some of the project's packages can be installed on any Ubuntu LTS machine, such as the packaged Jetbrains IDEs.

Maratona Linux also ships with the recommended IDEs unavailable on Ubuntu's repositories: Microsoft's proprietary build of their Visual Studio Code; Jetbrains's proprietary CLion; and the community editions from Jetbrain's PyCharm and IntelliJ.

Other notable differences with the official ICPC disk are the addition of dictionary packages, which are important as many contestants are not native English speakers; and the availability of the source code and the development documentation of the custom tools developed for the project.

Maratona Linux has already been successfully migrated from Ubuntu 18.04 to 20.04 \cite{davi_maratona_ubuntu_20}. This migration was responsible for replacing Snap-packaged IDEs with their Flatpak-based versions, modernizing the packaging tools to the version recommended by the operating system, and packaging the Visual Studio Code extensions instead of trying to download them \cite{davi_maratona_ubuntu_20}.

\subsection{Debian Packages}
Debian packages are used by over 100 Debian GNU/Linux-derived distributions, such as Ubuntu, to ship software on multiple processor architectures and system kernels \cite{Nussbaum2021}. They follow the Debian binary package format, which consists of a compressed standardized directory structure containing control files and the data to be installed \cite{deb_manpage_bullseye}.

The files' syntax, the requirements regarding packaging details and dependency resolution, and how the package should behave during the installation, removal and update procedures are regulated by the Debian Policy Manual \cite{debian_policy}. All official Deb packages are encouraged to observe the requirements stated on a specific version of the Policy Manual, preferably the most recent one.

In a Debian-derived operating system, Debian packages have the extension \texttt{.deb} and are managed and built using a dedicated low-level suite: \texttt{dpkg} \cite{dpkg_manpage_debian_bullseye}. System administrators will often manage the packages in their machines using the \texttt{apt} or \texttt{aptitude} interactive front-ends, and developers will often use the building tools and helper scripts, such as \texttt{debuild}.

Debian-derived distributions often distribute their packages on their repositories, which have a standardized format \cite{debian_repository_format}. Installing software unavailable in the distribution's repositories may be achieved by configuring an external repository. Ubuntu-based distributions automate this procedure by allowing packages built on Canonical's Launchpad platform to be added using short addresses \cite{launchpad_ppa}.

\subsection{Snap Packages}
The Snap packaging ecosystem is developed by Canonical Ltd., which is also responsible for the development of the Ubuntu Linux distribution. It is supported by other Linux distributions, yet it is more common on Ubuntu and its derivatives. Although the ``snap'' format; the management application, also called ``snap''; the daemon program \texttt{snapd}; and the packaging framework \texttt{snapcraft} are open-sourced; the main repository, maintained by Canonical, Snap Store, remains proprietary.

Snap packages have their dependencies and expected directory structure bundled in a compressed read-only filesystem (SquashFS) and may be run inside a sandbox \cite{canonical_snap_glossary_2022}. The \texttt{snapd} daemon mounts the compressed images, launches the required applications, and performs package management tasks requested by the system's administrator.

In order to avoid excessive dependency duplication, developers are encouraged to package their applications specifying ``base snaps'', which are a set of common libraries and runtimes currently denominated ``cores'' and are based on Ubuntu LTS releases \cite{canonical_snap_glossary_2022}. This containerized approach may still cause unnecessary library duplication in the user's machine when applications require different versions of a library included in a ``base snap''.

\subsection{Flatpak Packages}
Flatpak packages are similar to the Snap ones in their dependency bundling and sandboxing features. The main differences rely on the technologies used to achieve these functionalities, as the Flatpak framework is focused on the GNU/Linux desktop platform and requires the following of some Freedesktop conventions in order to integrate correctly with the user's desktop environment \cite{flatpak_under_the_hood_2022}.

Unlike the Snap packaging ecosystem, the Flatpak framework does not require running a daemon to perform common tasks or to run applications, yet programs can depend on daemons commonly found on GNU/Linux desktop environments, e.g. the D-Bus message bus system \cite{flatpak_under_the_hood_2022}. Packages can also be hosted on any online repository. Developers often choose Flathub, an online repository that also offers an infrastructure to build packages.

Flatpak packages, also called bundles, use the OSTree format to reduce the disk overhead used when multiple ``common packages'' are required by the user's packages. It uses a Git-like approach to track binary files: packages are pulled from a branch on a remote repository to a local one and identical files are stored only once \cite{flatpak_under_the_hood_2022}.
	
	\section{Development and Tests}
The system was developed in an AMD64 computer with Debian GNU/Linux version 12.0, codenamed Bookworm/testing. The development environment consisted of tools for generating and checking Debian packages and running a virtual machine used to test its installation. They are available in Table \ref{tab:program_versions}. The project’s source code and bug reporting for developers are available on Git repositories in the ``maratona-linux'' organization in Github \footnote{\url{https://github.com/maratona-linux}}. A Launchpad team \footnote{\url{https://launchpad.net/~icpc-latam}} organizes the repositories for the binary packages in their development \footnote{\url{https://launchpad.net/~icpc-latam/+archive/ubuntu/unstable}} and stable versions \footnote{\url{https://launchpad.net/~icpc-latam/+archive/ubuntu/maratona-linux}}.

\begin{table}[ht]
	\centering
	\caption{Packages and their versions used in the migration process of Maratona Linux}
	\begin{tabular}{@{}lll@{}}
		\toprule
		\textbf{Package}   & \textbf{Version} & \textbf{Programs}     \\ \midrule
		devscripts         & 2.23.2           & debclean, debuild     \\
		lintian            & 2.116.3          & lintian               \\
		shellcheck         & 0.9.0-1          & shellcheck            \\
		git-buildpackage   & 0.9.30           & gbp-dch               \\
		dpkg-dev           & 1.21.21          & dpkg-scanpackages     \\
		debhelper          & 13.11.4          & -                     \\
		debmake            & 4.4.0-1          & debmake               \\
		piuparts           & 1.1.7            & piuparts              \\
		desktop-file-utils & 0.26-1           & desktop-file-validate \\
		libguestfs-tools   & 1:1.48.6-2       & guestfish             \\
		qemu-system-x86    & 1:7.2+dfsg-5     & qemu-system-x86       \\
		qemu-utils         & 1:7.2+dfsg-5     & qemu-img              \\
		ovmf               & 2022.11-6        & -                     \\
		build-essential    & 12.9             & -                     \\
		gpg                & 2.2.40-1.1       & gpg                   \\
		\bottomrule
	\end{tabular}
	\label{tab:program_versions}
\end{table}

On each modification performed on the source code, the packages were rebuilt using a compilation script. New packages created to implement new features or to reduce the larger ones were created based on the base package created by the tool \texttt{debmake}.

Shell scripts in the source code were statically analyzed by \texttt{shellcheck}. Graphical shortcuts, implemented in GNU/Linux as \texttt{free\-desktop.org} Desktop Entries, are analyzed statically by \texttt{desk\-top-file-va\-li\-da\-te}. When a new version had to be released, the \texttt{debian/changelog} files were updated using \texttt{gbp-dch} to generate a template. Binary packages were generated from the source files using the compilation script.

The compilation process uses the tools \texttt{debclean} to clean the build results; \texttt{debuild}, to generate Debian binary packages; \texttt{lintian}, to check packages for Debian Policy violations; \texttt{dpkg-scanpackages}, to generate an index of all packages required to build a custom repository; and a custom tool to generate a dependency graph. The \texttt{piuparts} tool is used to test the installation of packages that greatly modify the system.

Manual tests were run every time a new build was deployed on the local repository. They consisted of installing the system on an updated Ubuntu 22.04 LTS virtual machine on \texttt{qemu-system-x86\_64} with the OVMF UEFI firmware. A secondary disk with the binary packages and the package index was set up as a local repository using \texttt{fstab} for automatic mounting and \texttt{/etc/sources.list} for trusting a local non-signed repository. Shell scripts partially automate the VM initialization and execution.

All IDEs (Visual Studio Code, IntelliJ, PyCharm, Clion, Codeblocks) and editors (Emacs, Vim, Gedit) and all included programming languages (C, C++, Python, Java and Kotlin) are tested after the installation of the \texttt{maratona-desktop} meta-package, responsible for installing the most common packages in a contest setup.

After the local tests were successful, the packages would be published on the Launchpad repository. This requires the developer to upload the packages compiled in the source format and signed by an OpenPGP key. The Launchpad platform compiles the source packages into binary ones and makes them available on a public repository. These packages are then tested on the virtual machine using the same procedures used to test the locally compiled binary packages.
	
	\section{Migration Procedures}
Maratona Linux was successfully migrated from Ubuntu LTS 20.04 to 22.04 \cite{davi_maratona_ubuntu_20}. This process required upgrading the package build tools and removing deprecated or incompatible programs and configurations. The development process was also improved by the introduction of an installation test for the binary packages and tools for creating packages that comply with the Debian Policy; the division of the largest packages into smaller ones; and the creation of new packages to fix time synchronization problems.

\subsection{Changes in the development environment}
All the development tools except \texttt{debhelper} were upgraded since the last migration to Ubuntu 20.04. There were no significant incompatibilities on most tools, yet the upgrade to QEMU 7.0 broke the virtual machine setup due to a lack of component versioning.

The \texttt{debhelper} compatibility level was increased from 12 to 13, following the recommended packaging environment for Ubuntu 22.04 \cite{debhelper_manpage_ubuntu_jammy}. This change was performed on the main configuration file, \texttt{debian/control}, of each package.

Two tools were introduced to the development pipeline: \texttt{debmake} and \texttt{piuparts}. The former was used to generate the templates that enabled the new packages, \texttt{ma\-ra\-to\-na-kai\-ros} and \texttt{ma\-ra\-to\-na-u\-su\-a\-rio-icpc}, to be compliant with the Debian Policy 4.6.0.1. The latter was used to test whether the package that changed the operating system the most, \texttt{ma\-ra\-to\-na-u\-su\-a\-rio-icpc}, had installation problems.

\subsection{Documenting the virtual machine configuration}
The virtual machine (VM) configuration and execution are now documented and centralized in a Github repository. These scripts also specify the version of the system's components instead of relying on the default ones, which may change in future QEMU releases. Versioning also enables a seamless migration of the guest machines, as new releases are expected to support the machine types from previous ones \cite{qemu_manpage_bullseye}.

The scripts now explicitly declare the machine type version as a KVM-accelerated \texttt{pc-q35-7.0}, and the CPU model as an Intel Haswell. This also helps to standardize the virtual machine's architecture, as the default CPU, \texttt{qemu64}, is subject to changes between releases and is often set to emulate an old model vulnerable to Spectre attacks \cite{qemu_cpu}.

The scripts also standardize the guest system's network devices, graphical processor, storage devices and bus connections. The VM connects its peripherals through a PCI bus: the SCSI-based hard disk drives, the network adapter and the display device.

\subsection{Creating the new packages}
The package \texttt{ma\-ra\-to\-na-kai\-ros} fixes synchronization problems with the Brazilian Legal Time (Hora Legal Brasileira). It replaces the standard NTP client, \texttt{systemd-timesyncd}, with \texttt{chrony}. This replacement enables synchronization even on unstable Internet connections. From \texttt{ma\-ra\-to\-na-me\-ta}, \texttt{ma\-ra\-to\-na-u\-su\-a\-rio-icpc} was extracted to centralize contestants' user creation and to reduce the size and complexity of a bigger package.

\subsection{Upgrading the external IDEs and removing the dependency on Flatpak}
Maratona Linux used to install the programs unavailable on Ubuntu's repositories using the Flatpak framework and the Flathub repository. Unfortunately, the sandboxing approach's limitations negatively impacted the project's users: they could not compile C and C++ code or read manpages in the Visual Studio Code's terminal. Also, fixing differences between compilers and Glibc versions in the system and the runtime required several workarounds that had to be revised on each release.

The external IDEs were downloaded by the package \texttt{ma\-ra\-to\-na-e\-di\-to\-res-flat\-pak}, which specified their versions and dependencies using their OSTree commit hashes. This approach was replaced by native Debian packages that install the aforementioned IDEs according to the versions in Table \ref{tab:ext-ides-versions}.

\begin{table}[htb]
	\centering
	\caption{External IDEs and their versions.}
	\begin{tabular}{@{}lll@{}}
		\toprule
		\textbf{IDE}            & \textbf{Package}                     & \textbf{Version} \\ \midrule
		PyCharm Community       & \texttt{maratona-intellij-pycharm}   & 2022.1.4         \\
		IntelliJ IDEA Community & \texttt{maratona-intellij-idea}      & 2022.1.4         \\
		CLion                   & \texttt{maratona-intellij-clion}     & 2022.1.3         \\ 
		Visual Studio Code      & \texttt{maratona-visual-studio-code} & 1.73.1           \\
		\bottomrule
	\end{tabular}
	\label{tab:ext-ides-versions}
\end{table}

\begin{table}[htb]
	\centering
	\caption{Visual Studio Code's required extensions for C, C++, Java, Kotlin and Python3 development}
	\begin{tabular}{@{}ll@{}}
		\toprule
		\textbf{Extension}        & \textbf{Version}  \\ \midrule
		vscode-cpptools           & 1.14.3            \\
		formulahendry.code-runner & 0.12.0            \\
		vscjava.vscode-java-pack  & 0.25.7            \\
		fwcd.kotlin               & 0.2.26            \\
		ms-python.python          & 2022.18.2         \\ \bottomrule
	\end{tabular}
	\label{tab:vscode_extensions}
\end{table}

The \texttt{ma\-ra\-to\-na-vscode-extensions} installs the required extensions, available in Table \ref{tab:vscode_extensions}, to enable C++, Java, Kotlin and Python3 support in Visual Studio Code. It uses the same approach found on ICPC's World Final disk: only the contents of the \texttt{extensions} directory inside the \texttt{vsix} package are extracted to a dedicated directory.

\subsection{Tracking modifications}
\begin{figure}[htb]
	\centering
	\begin{overpic}[abs, unit=1mm, height=0.66\textheight, trim= 1mm 0 1mm 0, clip]{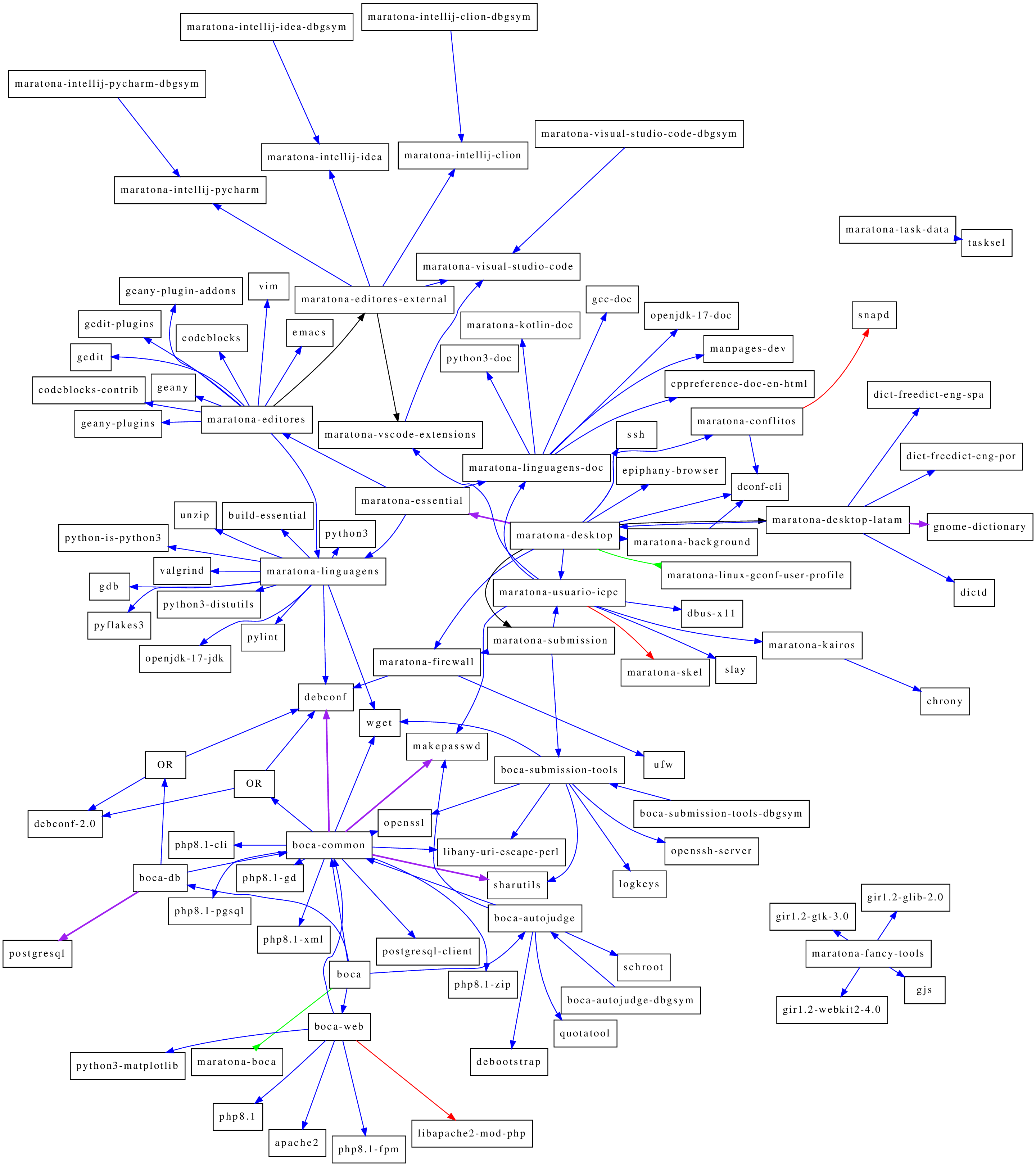}
		\put(90,137){\includegraphics[width=0.15\linewidth, trim=45mm 90mm 30mm 0, clip]{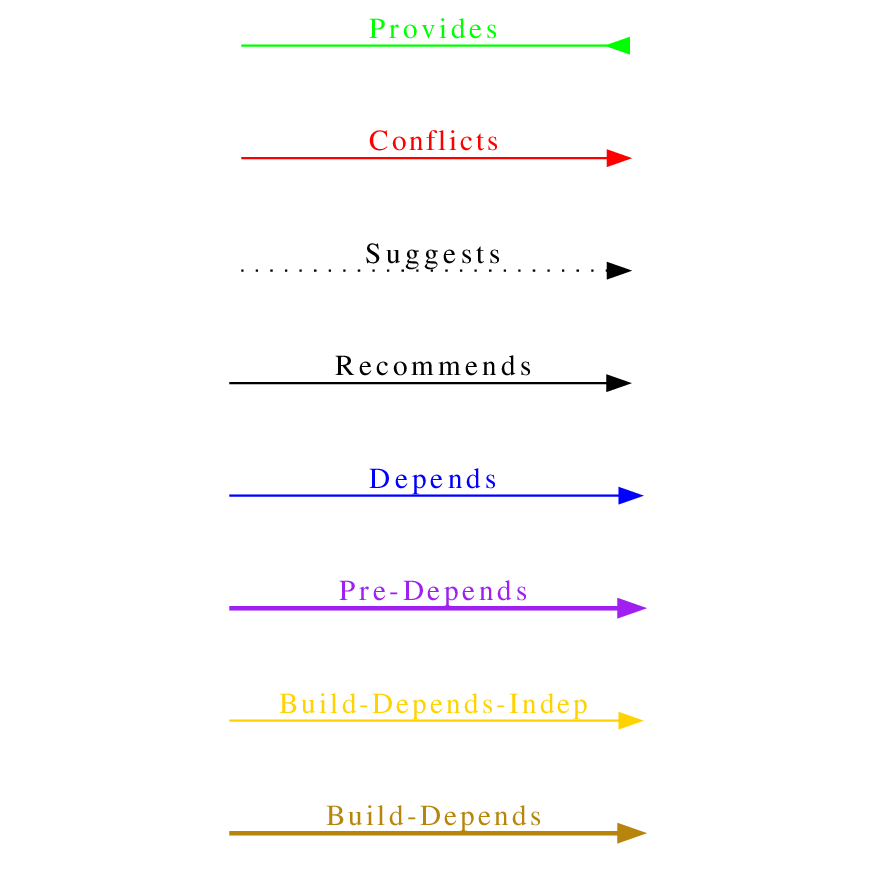}}
		\put(113,127){\includegraphics[width=0.15\linewidth, trim=45mm 0 30mm 50mm, clip]{imgs/legend.eps}}
	\end{overpic}
	\caption{Dependencies of Maratona Linux's packages and BOCA online judge}
	\label{fig:dependencies}
\end{figure}

Modifications made to the packages were logged in Git commit messages and the package's \texttt{control}, \texttt{changelog}, and \texttt{copyright} files. The project's dependency graph, available in Figure \ref{fig:dependencies}, was updated with the new packages. The graph is generated by a script which uses the official Debian Python bindings to analyze a local repository and map the dependencies between the packages.


The script generates a dependency graph using the conventions of another Debian package dependency tracker, \texttt{debtree}. The former can generate a graph from a local repository, unlike the latter. The edges define the multiple types of dependencies available to a package, according to the Debian Policy Manual \cite{debian_policy}. The most prevalent blue edges represent the most basic level, \texttt{Depends}, in which a package ``absolutely depends'' on another to provide functionality, requiring the dependency's installation \cite{debian_policy}.

Black solid edges represent a \texttt{Recommends} dependency, in which a package strongly recommends the installation of its dependency \cite{debian_policy}. The purple edges, on the other hand, determine an even stricter dependency relation than the previously mentioned blue edges: the \texttt{Pre-Depends} relationship. It ought to be used only when strictly necessary, and requires that the dependency package must be fully installed before the dependent package starts its installation procedures \cite{debian_policy}.

Other relationship types present in figure \ref{fig:dependencies} are the \texttt{Pro\-vi\-des}, represented by the green edge with an inverted arrow, and the \texttt{Conflicts}, pictured by a red edge. The former allows the definition of virtual packages, and the latter declares a hard incompatibility \cite{debian_policy}.

\subsection{Migrations from Snap to a native Debian package}
Canonical migrated the Firefox web browser from a native Debian package to a Snap-based one. This sharply increased the difficulty of the customization of the launch options necessary to facilitate contestants' actions during the competition. In order to remain with the custom launch options, the default web browser was replaced by GNOME Web, also known as Epiphany, which is packaged as a native Debian package.

Compared to Firefox, Epiphany supports fewer web standards \cite{fedora_magazine_browsers}. This does not significantly impact Maratona Linux, as the web pages used for submitting the solutions do not use advanced features. Furthermore, Epiphany exposes many of its settings as keys in the \texttt{dconf} database, which can be easily modified with text files.

\subsection{Improved compliance with the Debian Policies}
Although the new packages are compliant with the Debian Policy version 4.6.0.1 and progress was made regarding the compliance of older packages, the latter still have policy violations detected by \texttt{lintian}. Most of the fixed violations were related to packages with missing dependencies, and control files that did not comply with the syntax or the best practices.


The older packages still need modifications to further comply with the policy: implement maintainer scripts to uninstall or upgrade packages; migrate packages to the current standard, \texttt{3.0}; reduce the complexity of the larger packages to speed the static analysis up; fix the policy violations regarding \texttt{debconf}; unify all the modifications in the system's \texttt{dconf} database; adopt the recommended practices regarding \texttt{systemd} services.
	
	
	\section{Final Remarks}
Maratona Linux was successfully migrated from Ubuntu LTS version 20.04 to 22.04. Its maintainability was improved by unifying package building tools; introducing static analyzers; tracking the project's modifications; listing the developers' contacts; and updating changelogs, descriptions, and licenses. Packaging the external IDEs as Debian packages enabled the teams to reliably use embedded terminals and the system's default compilers and debuggers, greatly improving the system's usability during the competition.

Even though the improvements made, there are still opportunities for enhancing the project. Ubuntu LTS version 24.04 will soon be released. This may require packages or the packaging pipeline to be updated due to dependency or Debian Policy changes. There are still some deviations from the Debian Policy that can be amended.

Among the remaining violations of the current Debian Policy are: the lack of scripts for reinstalling, removing and updating; the interactive installation system using scripts that are not compliant with the restrictions imposed by the Policy; and some modifications in the system behavior are not being performed using the recommended tools.
	
	\section*{Acknowledgments}
	This research is part of the University of Brasilia's (UnB) technological development and innovation-focused scientific initiation projects PIBITI 2020-2021 and PIBITI 2021-2022. The authors kindly thank UnB for their financial support.
	
	\printbibliography
	
	
	
\end{document}